\documentclass[aps,prl,preprint,superscriptaddress]{revtex4-1}
\usepackage{graphicx}
\usepackage{gensymb}
\usepackage{color}
\usepackage{amsfonts}
\usepackage{isomath}
\usepackage{amsmath}
\usepackage{amsthm}
\usepackage{amsfonts}
\usepackage{braket}
\usepackage{siunitx}

\usepackage{yhmath}

\def\gp{${\mathrm{\Gamma}}$}

\begin{document}

\title{Spectroscopic signatures of persistent exciton condensation in a bulk magnetic topological insulator}

\author{Paulina Majchrzak}
\email{pemaj@stanford.edu}
\affiliation{Department of Applied Physics, Stanford University, Stanford, California 94305, USA}
\author{Chakradhar Sahoo}
\affiliation{Department of Physics and Astronomy, Aarhus University, 8000 Aarhus C, Denmark}
\author{Manuel Tuniz}
\affiliation{Dipartimento di Fisica, Università degli Studi di Trieste, 34127 Trieste, Italy}
\author{Wibke Bronsch}
\affiliation{Elettra - Sincrotrone Trieste S.C.p.A., 34149 Basovizza, Italy}
\author{Denny Puntel}
\affiliation{Dipartimento di Fisica, Università degli Studi di Trieste, 34127 Trieste, Italy}
\author{Federico Cilento}
\affiliation{Elettra - Sincrotrone Trieste S.C.p.A., 34149 Basovizza, Italy}
\author{Xing-Chen Pan}
\affiliation{Advanced Institute for Materials Research, Tohoku University, Sendai 980-8577, Japan}
\author{Jakob Kjærulff Svaneborg}
\affiliation{Department of Physics, Technical University of Denmark, DK-2800 Kongens Lyngby, Denmark}
\author{Yong P.  Chen}
\affiliation{Department of Physics and Astronomy, Aarhus University, 8000 Aarhus C, Denmark}
\affiliation{Advanced Institute for Materials Research, Tohoku University, Sendai 980-8577, Japan}
\author{S{\o}ren Ulstrup}
\email{ulstrup@phys.au.dk}
\affiliation{Department of Physics and Astronomy, Aarhus University, 8000 Aarhus C, Denmark}

\begin{abstract}

Exciton condensates are long-sought correlated quantum states arising from macroscopic coherence of bound electron-hole pairs. Although equilibrium and transient excitonic states have been reported in several material platforms, direct evidence for a light-induced exciton condensate state has been challenging to achieve as an intrinsic property of a bulk quantum material. Here, we use time- and angle-resolved photoemission spectroscopy to investigate long-lived photoexcited carriers in the intrinsic magnetic topological insulator MnBi$_2$Te$_4$ with the chemical potential tuned to the topological surface state by Sb substitution. Following optical excitation, we observe the delayed emergence of a transient state whose formation coincides with depopulation of the bulk conduction band and whose lifetime extends to the microsecond timescale. The quasiparticle dispersion exhibits a pronounced flattening and develops a Mexican-hat-like profile. These spectral signatures are consistent with the formation of a metastable excitonic condensate state. Our results establish magnetic topological insulators as a promising platform for investigating long-lived photoinduced many-body states and their interplay with topology and magnetism.

\end{abstract}

\maketitle

\section{Introduction}

Excitons, bosonic excitations composed of bound electron-hole pairs, shape the optical responses of semiconductors and provide a rich playground for realizing novel quantum phenomena \cite{Wilson:2021, Wu:2024}. Amongst those, an exciton condensate (EC) is highly sought after as it creates macroscopic phase coherence and introduces strong Coloumb correlations. The behavior of excitons in a condensate is highly tuneable via the exciton density. In the weakly interacting, dense limit, excitons behave as Cooper pairs and follow the Bardeen–Cooper–Schrieffer (BCS) formalism such that the excitation spectrum departs from the single-particle dispersion via gap-opening and band renormalization effects \cite{Perfetto:2019,Mori:2025}.
 
The equilibrium EC phase has been previously reported in charge-density wave systems \cite{Kogar:2017}, graphene \cite{Li:2017, Rickhaus2021} as well as monolayers \cite{Sun:2022, Jia:2022} and two-dimensional (2D) heterostructures of transition metal dichalcogenides \cite{Qi:2026}.  However, owing to the charge neutrality of these correlated excitations, unambiguous demonstration of light-induced condensation is difficult by optical methods. Recently, momentum-resolved spectroscopy was predicted \cite{Rustagi2018, Christiansen2019} and then experimentally confirmed \cite{Madeo:2020, Man2021,Schmitt:2022} to enable identification of a non-equilibrium exciton population, providing an ideal probe of the quasiparticle excitation spectrum that potentially gives access to a light-induced condensate phase.

MnBi$_2$Te$_4$, the first experimentally realized intrinsic magnetic topological insulator, exhibits several characteristics making it a promising candidate for hosting an EC phase. It has been suggested that a transient gap can be stabilized in the topological surface state (TSS) of a topological insulator, resulting in band flattening and appearance of sidebands \cite{Triola:2017, Pertsova:2018}.
Moreover, the interplay of non-trivial topology and magnetic order in MnBi$_2$Te$_4$ makes for a platform associated with several exotic phenomena, including the quantum anomalous Hall effect (QAHE) \cite{Yujun2020} and an axion insulating state \cite{Liu2020}. The  spin structure and topology of the EC state would be governed by the competition between Coulomb interactions and magnetism \cite{Liebman:2025, Stamper:2013}.

In bulk crystals of MnBi$_2$Te$_4$, high levels of manganese vacancies typically result in substantial $n$-doping \cite{Du:2021}. The intrinsic occupation of the bulk conduction band presents challenges for realizing electron transport through the TSS and may affect the material's properties, including its optical response. Substitution of bismuth atoms with antimony enables efficient tuning of the chemical potential, suppressing the bulk carrier conductivity while preserving the topologically non-trivial character and spontaneous magnetization \cite{Chen:2019b, Yan2019_mbst}.

Here, we employed time- and angle-resolved photoemission spectroscopy (trARPES) to characterize the non-equilibrium electronic structure of Sb-doped MnBi$_2$Te$_4$, where the chemical potential was tuned into the TSS. The transient state suggests formation of excitons with a dispersion that is consistent with a BCS-like excitonic state: Its formation exhibits a picosecond delay due to the depopulation of the bulk conduction band, it exhibits a lifetime on the microsecond timescale and the exciton dispersion resembles a Mexican-hat shape.

\section{Results and Discussion}

\begin{figure*}[t!] 
\begin{center}
\includegraphics[width=\textwidth]{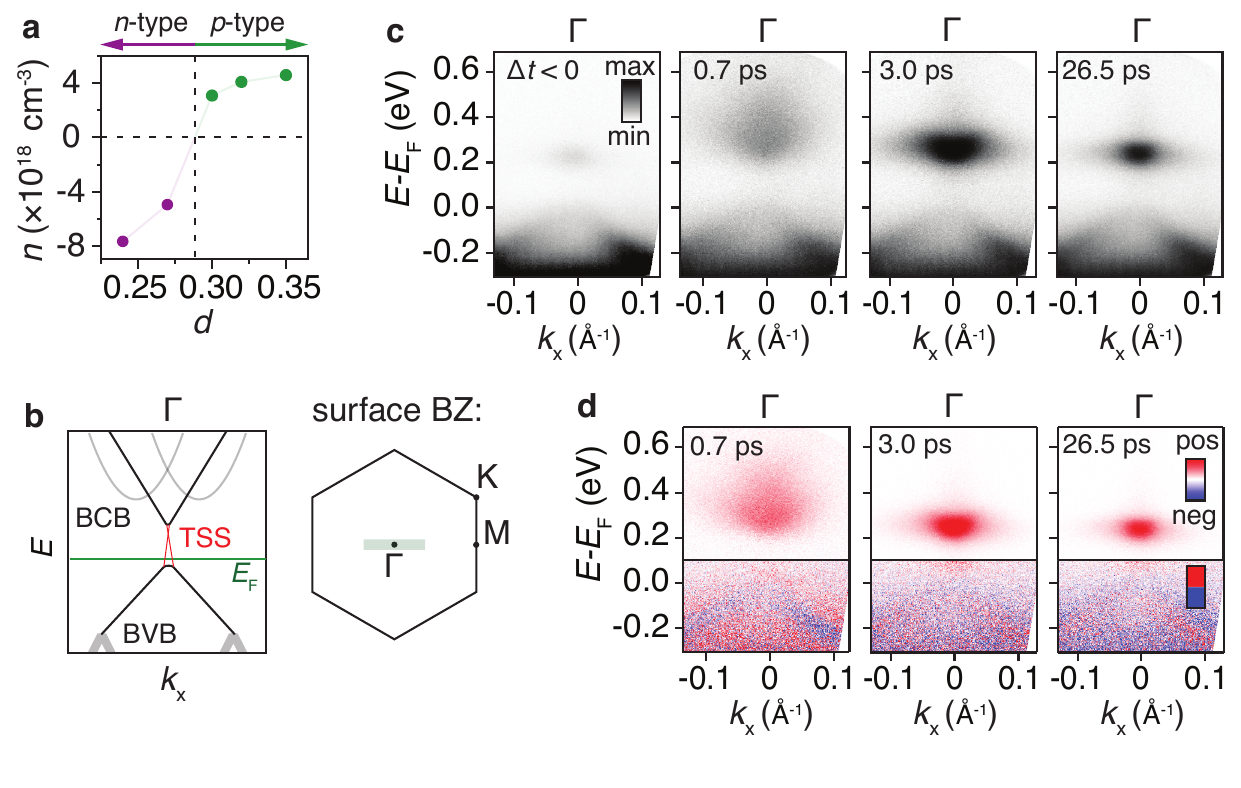}
\caption{\textbf{Chemical and optical doping of a magnetic topological insulator.} (a) Carrier concentration, derived from Hall resistivity measurements, as a function of nominal Sb mole fraction $d$ in MnBi$_{2(1-d)}$Sb$_{2d}$Te$_4$ (MBST). Dashed vertical line marks the transition from $n$- to $p$-doping. (b) Schematic band structure of MBST and the surface Brilloin zone (BZ). Bulk valence band (BVB), bulk conduction band (BCB) and topological surface state (TSS) are marked on the diagram. Green solid line indicates the position of the equilibrium chemical potential in the crystal considered here, with $d = 0.3$. Green shaded line indicates the cut direction across the BZ. (c) Photoemission spectra of MBST taken at selected time delays. Note that the averaged above-Fermi level signal before optical excitation ($\mathrm{\Delta} t < 0$) is a result of the sample not relaxing fully between pump pulses arriving every 4~\textmu s. (d) Corresponding intensity difference spectra obtained by subtracting the average spectrum measured for $\mathrm{\Delta} t < 0$ from the excited state signal at the given time delays. Note that the bottom part of the spectra (below the solid line) uses a saturated color scale (see the extra color bar).}
\label{fig:1}
\end{center}
\end{figure*}

Figure~\ref{fig:1}(a) shows the carrier density extracted from Hall resistivity measurements in the MnBi$_{2(1-d)}$Sb$_{2d}$Te$_4$ (MBST) series. In agreement with previous studies \cite{Chen:2019b}, for a nominal Sb mole fraction of $d = 0.28$, the majority carrier character changes from electrons ($n$-doping) to holes ($p$-doping). In Fig.~\ref{fig:1}(b), the band structure of MBST is sketched out, based on our previous high-resolution ARPES measurements \cite{Volckaert:2023}. At substitution fraction $d = 0.3$, investigated here, the chemical potential is placed in the bulk band gap, where the TSS is located, close to the V-shaped bulk valence band (BVB) edge. This is confirmed by our trARPES measurements presented in Fig.~\ref{fig:1}(c), using 6.2~eV probe pulses. With 1.55~eV pump pulses at a fluence of 60~\textmu J~cm$^{-2}$, we excite electrons well above the band gap, such that they subsequently occupy the bulk conduction band (BCB) up to 0.6 eV above the equilibrium Fermi level. The excited electrons then undergo cascading interlayer scattering \cite{Sobota2012, Majchrzak:2023} and accumulate at the bottom of the BCB after 3.0~ps. A considerable fraction ($\sim70\%$) of carriers remains excited after at least 26.5~ps. To further visualize the transient signal, we plot the difference between the spectrum taken at a given time delay $\mathrm{\Delta} t$, and the average of spectra collected before the arrival of the optical excitation ($\mathrm{\Delta} t < 0$) in Fig. \ref{fig:1}(d). The red (blue) signal corresponds to gain (loss) of the photoemission intensity. Note that before optical excitation, $\mathrm{\Delta} t < 0$, a faint signal ($\sim8\%$ of the intensity present at 3.0~ps) is present above the Fermi level, which bears a strong resemblance to the transient signal we observe at long time delays after excitation. This signal is not present in our equilibrium ARPES measurements without optical excitation. This observation suggests that the system may not be fully relaxed in the time between two consecutive pulses (at 250~kHz, this corresponds to 4~\textmu s), therefore the spectrum at $\mathrm{\Delta} t < 0$ does not represent a true equilibrium. We show the static characterization in Supplementary Note~1.

\begin{figure*}[t!] 
\begin{center}
\includegraphics[width=\textwidth]{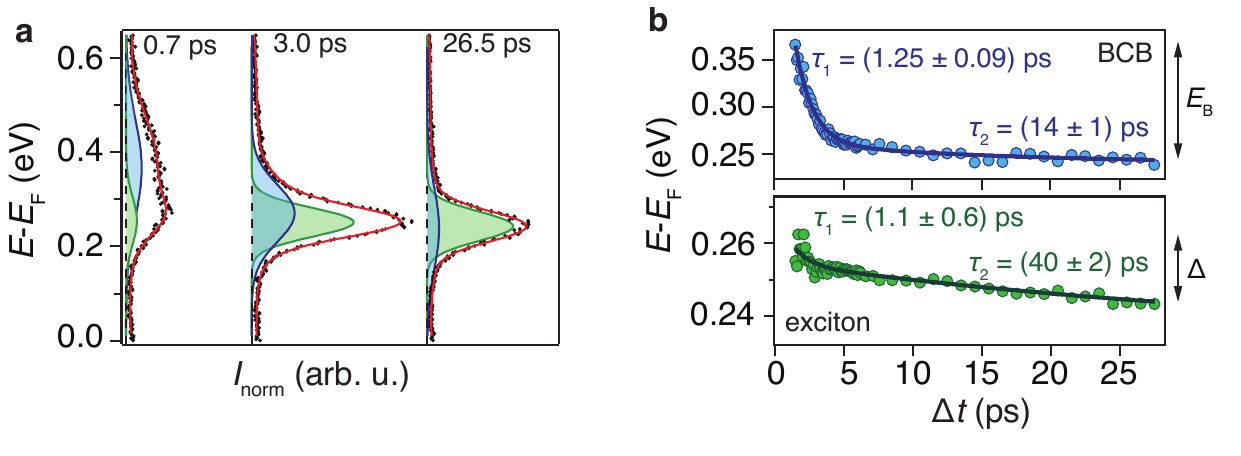}
\caption{\textbf{Overview of dynamics at $T = 110$~K.} (a) Energy-distribution curves (EDCs) around the \gp-point at selected time delays (black markers) fitted with a double Lorentzian function (red curve). Peak components are shaded in blue and green. (b) Positions of the fitted peaks as a function of time, with double exponential fits overlayed. The overall shift of the upper and lower peak provide estimates for the exciton binding energy, $E_\mathrm{B}$, and the BCS gap, $\mathrm{\Delta}$, respectively. The magnitudes of $E_\mathrm{B}$ and $\mathrm{\Delta}$ are demarcated by double-headed arrows.}
\label{fig:2}
\end{center}
\end{figure*}

To elucidate the origin of the exceptionally long-lived transient signal, we examine how the energy-distribution curves (EDCs) around the \gp-point change with time delay after optical excitation in Fig.~\ref{fig:2}(a). The excited state population is initially well-described by two Lorentzian peaks on a linear background. The higher-energy peak (in blue) is broad, reflecting the distribution of carriers across the ladder of BCB states. For the first few picoseconds after the pump arrival, its center position gradually decreases. By contrast, the  position of the lower-energy peak remains almost constant, while its amplitude increases, and its linewidth sharpens. Eventually, the EDCs at later delays can be described exclusively by a single peak. In the following, we will argue that the bottom peak at long delays is suggestive of an excitonic energy level that undergoes gap opening as a result of condensation of electron-hole pairs into a BCS state.
In Fig.~\ref{fig:2}(b) we track the position of the two peaks as a function of time. At $\mathrm{\Delta} t = 0.7$~ps, the peaks are located 100~meV apart, but their separation exponentially diminishes. The transient shift of both peaks can be described by a double exponential decay. The shorter timescale, $\tau_1 \approx 1.1 - 1.2$~ps, describes the time required for carriers to transfer from the BCB to the exciton state and form the bound electron-hole pairs. The longer timescale, $\tau_2$, which reaches $\approx 40$~ps in the exciton state reflects the timescale required to attain the metastable EC phase.

\begin{figure*}[t] 
\begin{center}
\includegraphics[width=\textwidth]{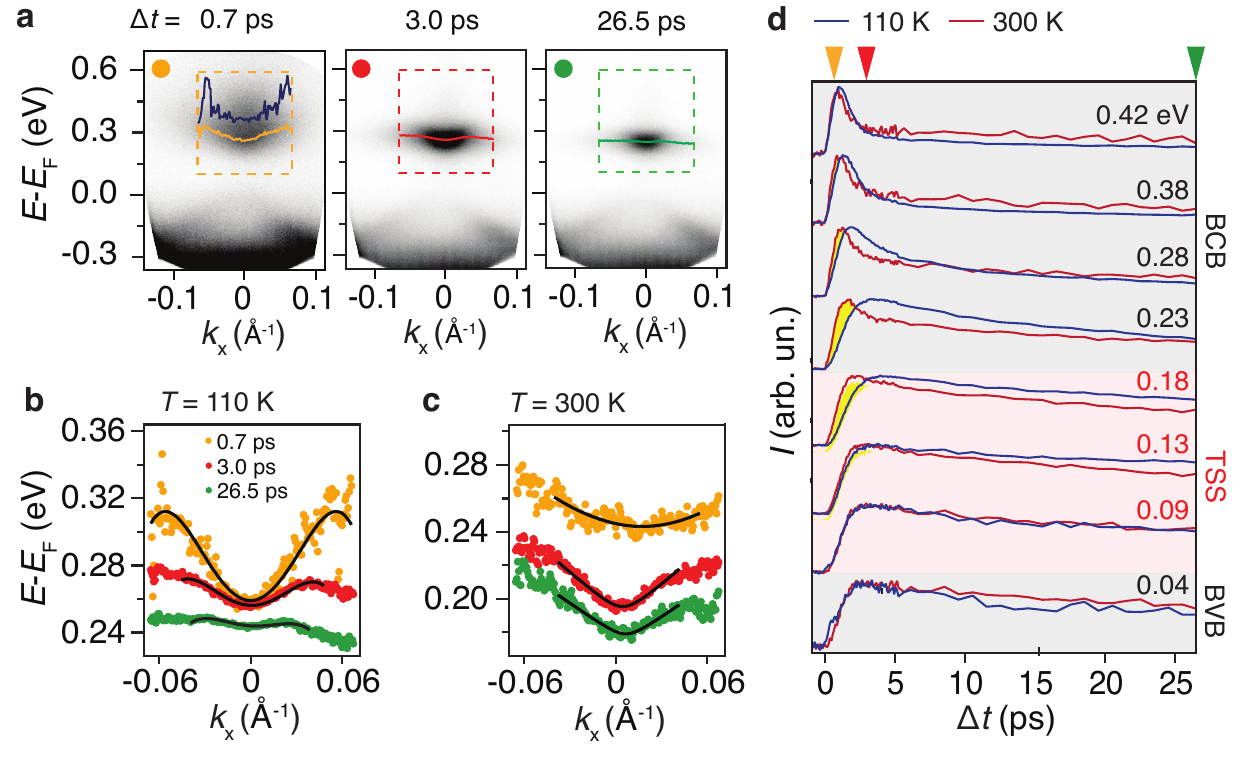}
\caption{\textbf{Evolution of band dispersion with time and temperature.} (a) ARPES spectra at selected time delays $\mathrm{\Delta} t$, with conduction band (dark blue curve) and exciton (yellow, red and green curves) dispersions extracted through EDC fitting presented in Fig.~\ref{fig:2}. The measurement was performed at $T = 110$~K. (b) Evolution of exciton dispersion as a function of time delay (color-coded as in (a)) at $T = 110$~K. Black curves represent fits to a Mexican-hat dispersion on a linear background. (c) Corresponding dispersion measured at room temperature. Black curves represent fits to a hyperbolic function. (d) Time-dependent ARPES intensity integrated over (48~meV, 0.02~Å$^{-1}$)-wide boxes around the \gp-point, with central energies indicated in eV. The grey- and red-shaded background indicates whether the energy is within the BCB, BVB or TSS dispersion. Blue (red) curves correspond to a sample temperature of 110~(300)~K. The yellow shading highlights the difference in buildup time between 110~K and 300~K measurements. Color-coded arrows at the top demarcate the time delays selected in (a)-(c).}
\label{fig:3}
\end{center}
\end{figure*}

Next, we analyze the dispersion of the long-lived transient state. In Figure~\ref{fig:3}(a), results of EDC fits are overlayed on the trARPES spectra. For time delays after $\mathrm{\Delta} t = 0.7$~ps, a single peak is sufficient to reproduce the data, as shown in Supplementary Note~2. Plotting the lower energy level as a function of momentum for selected time delays together in Fig.~\ref{fig:3}(b) reveals that the dispersion becomes substantially flattened, transforming from an electron-like parabolic state to a Mexican-hat-like shape. This shape is usually associated with Bogoliubov quasiparticles and therefore suggestive of a BCS-type exciton condensate state \cite{Triola:2017}. In Supplementary Note~3, we show that the BVB is also modified at long delays, providing additional evidence consistent with this interpretation. The modification of the dispersion is not related to formation of transient magnetic order, for example through opening of a gap in the TSS, as our measurement temperature of 110~K is much higher than the Néel temperature of 25~K \cite{otrokov2019prediction}. The long lifetime of the effect also precludes an exciton-driven Floquet effect \cite{Chan:2023_Floquet, Pareek:2024}, whereby the sidebands would only be observable during the temporal overlap of the pump and probe pulses.

Combining the results in Figs.~\ref{fig:2} and \ref{fig:3} allows us to define two energy scales that describe the observed light-induced many-body state. The exciton binding energy, $E_B \approx 100$~meV, is given by the difference between the BCB edge at $\mathrm{\Delta} t = 0.7$~ps (when the majority of hot carriers have scattered down to the bottom of the band), and the energy of the exciton state. We note that the BCB comprises a complicated manifold of states \cite{otrokov2019prediction, Estyunin2020, Yan2021, Volckaert:2023}, and the EDC fitting is likely capturing an averaged band position, making our estimated value of $E_B$ the upper bound. Moreover, we observe a larger band gap in MBST compared to MnBi$_2$Te$_4$ without any Sb-doping. This can be explained by a reduced screening of the gap in the semiconducting regime compared to the bulk metallic parent compound.

In a BCS-like condensation, the pairing of electrons and holes determines critical temperature, $T_c$ \cite{Chen:2024}, such that the excitons are formed and condensed simultaneously. Therefore, the displacement of the excitonic state by 20~meV down in energy, from an upward-curving parabola at $\mathrm{\Delta} t = 0.7$~ps to the Mexican-hat-like dispersion at long time delays may be interpreted as the second energy scale, the opening of gap $\mathrm{\Delta}$ in the excitonic energy level. To account for the finite temperature of our measurement, we recall that the gap size decreases asymptotically as the temperature tends to $T_c$ \cite{BCS:1957}. From this relation, we can estimate the pairing temperature in MBST as $T_c \approx 150$~K. Since the BCB is composed of multiple bands, the electronic temperature is not easily extracted. We note the possibility that during the relaxation process, the temperatures of electronic and lattice subsystems may settle at a higher value than the base temperature. However, since we observe signatures of condensation, we conclude that our $T_c$ estimation represents a lower bound.
This establishes MBST as a platform for realizing a metastable high-temperature EC phase. The inferred $T_c$ is on par with charge-density wave exciton insulator candidates \cite{Kogar:2017}, and orders of magnitude higher than in 2D electron-hole bilayers \cite{Qi:2026}.

We contrast the above observations with data collected at room temperature (RT), presented in Fig.~\ref{fig:3}(c). In this condition, the EDCs can be described by a single band at all time delays (see Supplementary Note~3). The band dispersion does not show any signs of flattening, but rather evolves from a parabola to a V-shape. This reflects the equilibrium dispersion observed in the parent MnBi$_2$Te$_4$, and confirms that the pairing of electrons and holes does not occur. However, the BCB edge is shifted by 25~meV downwards with time delay, with the shift well-described by a double-exponential decay with two timescales, $\tau_1^\mathrm{RT} = (0.84 \pm 0.04)$~ps and $\tau_2^\mathrm{RT} = (25.5 \pm 0.9)$~ps, as demonstrated in Supplementary Note~4. This is consistent with the formation of electrons and holes in the high-density regime, beyond the Mott transition, causing a large band gap renormalization \cite{Chernikov2015_inv, Chernikov2015, Dendzik2020}. Then, the faster decay constant can be related to the electron-phonon scattering-mediated relaxation down the BCB, similar to that in the $n$-doped parent material MnBi$_2$Te$_4$ \cite{Majchrzak:2023}.
Shown in Fig.~\ref{fig:3}(d) is the full time dependence of the photoemission intensity integrated over (48~meV, 0.02~Å$^{-1}$)-wide windows, centered at different energies, for both $T = 110$~K and $T = 300$~K. 
At the bottom of the BCB and in the TSS, the intensity accumulation across the BCB proceeds faster at high temperature, as highlighted by the yellow shading, and consistent with the timescales extracted in Fig.~\ref{fig:2}(b). If the system was governed by electron-phonon scattering in both conditions, then the rise time for the signal should not depend strongly on temperature \cite{Sobota2014}. 
Instead, we speculate that this difference emerges from the accumulation of quantum coherence in the transient state below the EC transition temperature \cite{Nardin:2009}.

\begin{figure*}[t] 
\begin{center}
\includegraphics[width=\textwidth]{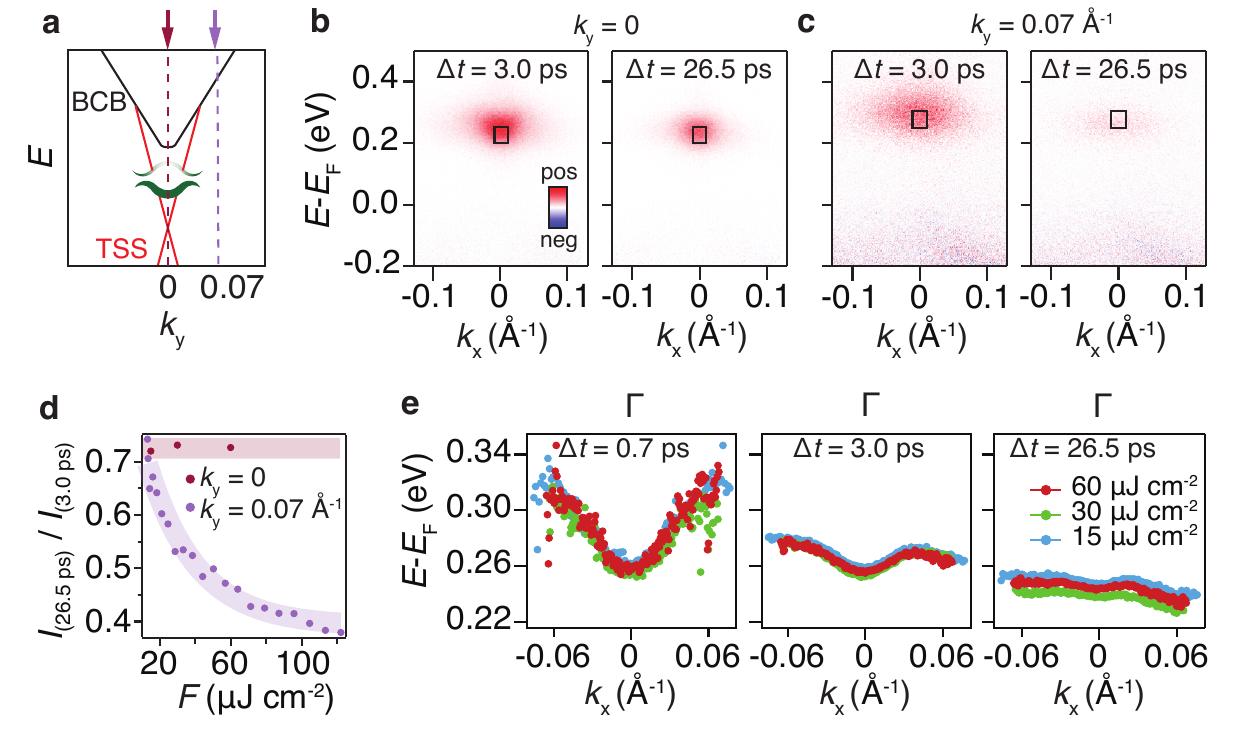}
\caption{\textbf{Fluence and momentum dependence of exciton state.} (a) Schematic illustrating the BCB and TSS dispersion with the EC phase. (b)-(c) Intensity difference spectra obtained at the stated $k_y$-values (see arrows and vertical dashed lines in (a)) at time delays corresponding to peak exciton population ($\mathrm{\Delta} t = 3.0$~ps) and long time delays ($\mathrm{\Delta} = 26.5$~ps). The spectra in (b) cut through the EC state while the spectra in (c) were obtained away from the EC bands. (d) Ratio of intensity integrated over the boxed regions in (b)-(c), extracted at the stated time delays and as a function of fluence $F$. Bold curves are provided as guides to the eye. (e) Extracted dispersions as a function of time delay and fluence. The same fitting procedure as in Fig.~\ref{fig:3}(b) was applied to obtain the dispersions.}
\label{fig:4}
\end{center}
\end{figure*}

In Figure~\ref{fig:4}, we consider the fluence dependence of the transient state.
To understand the behavior of the condensate state and the hot electrons in the BCB as a function of excitation density, we compare ARPES cuts around the \gp-point, and at a finite momentum offset from the \gp-point ($k_y = 0.07$~Å$^{-1}$), which does not contain the excitonic state, as shown schematically in Fig.~\ref{fig:4}(a). The difference spectra are presented in Figs.~\ref{fig:4}(b)-(c).
The ratio of the integrated intensity within the energy- and momentum-windows demarcated in panels (b)-(c) can provide information on the lifetime of the quasiparticles. This quantity is plotted in Fig.~\ref{fig:4}(d), revealing that around the \gp-point, the relative drop-off of intensity between early and late time delays stays constant with increasing fluence. On the other hand, for the BCB carriers at $k_y = 0.07$~Å$^{-1}$, the ratio of intensities decays exponentially with fluence.
To explain this discrepancy, we invoke the phenomenology related to Cooper pairs in the BCS theory. In the BCS EC, the electron-hole pairs are weakly interacting, but their density is very high, leading to significant wavefunction overlap. This overlap drives macroscopic coherence and causes the EC state to become highly resilient against scattering.  
Phenomenologically, this is described by the Rothwarf-Taylor bottleneck, whereby the density of bound pairs is large enough that their relaxation rate becomes limited by the probability of finding a scattering partner, saturating the dependence on excitation density \cite{Zhang_stimulated_2016}. By contrast, the inverse dependence of the BCB population on fluence is consistent with the bimolecular recombination of electron-hole pairs through the bulk bandgap, as previously seen in ultrafast reflectivity measurements, which are primarily bulk-sensitive, on an intrinsic topological insulator \cite{Gross:2021}. 
In Fig.~\ref{fig:4}(e), we additionally show the fitted dispersions of the excitonic state. At each delay, the dispersions for different fluences overlap with each other, demonstrating that the EC is robust within the investigated range of excitation densities.

Lastly, we discuss the conditions that allow for the build-up of coherence. In a conventional semiconductor, the optically induced excitons recombine and disappear before the coherent state can be stabilized. Here, we use excitation densities comparable to exciton-driven Floquet coupling in monolayer WS$_2$ \cite{Pareek:2024}. This suggests the presence of a mechanism counteracting the enhanced screening in our bulk material. A slow rate of electron-hole recombination can be achieved through spatial separation and/or symmetry-related bottlenecks. In our case of a magnetic topological insulator, the mixed dimensionality of the electronic states - 2D TSS and three-dimensional bulk bands - may allow for the formation of spatially indirect excitons with long lifetimes \cite{Mori2023, Mori:2025}. Moreover, the electron wavefunction may be dynamically redistributed between layers of the van der Waals material \cite{Lee2023}. Notably, a resonant excitation into an excitonic state is not required here, which further points to the key role of the TSS in preventing the electron-hole pairs from breaking up by a combination of topological protection \cite{Hajlaoui2014, Neupane:2015} and localization of the electron to the surface \cite{Mori2023}. As a result, the EC phase in our system becomes metastable.

\section{Conclusion}

In summary, we observed formation of a long-lived light-induced state that shows multiple hallmarks of BCS-like exciton condensation in an intrinsic magnetic topological insulator. This result is notable for several reasons. Firstly, it demonstrates that in topological materials, a non-equilibrium excitonic order may be generated even through a non-resonant excitation mechanism, providing a pathway to generate metastable coherence.
Secondly, with $T_c$ estimated to be $150$~K, our system represents a high-temperature counterpart to electrostatic ECs achieved in 2D materials at a milli- to few-kelvin range \cite{Li:2017, Rickhaus2021, Sun:2022, Jia:2022, Qi:2026}.
Finally, by applying chemical doping of an intrinsic magnetic topological insulator, we retain tuneability over the filling of the electronic bands, while not suppressing its non-trivial topological properties, which ultimately allows us to optically access the EC phase. The (MnBi$_2$Te$_4$)$_n$(Bi$_2$Te$_3$)$_m$ homologous family of magnetic topological insulators provides an especially fertile ground for exploring the intersection of many-body physics and topology, where multiple exotic phases can be accessed by inserting non-magnetic building blocks and varying the number of layers in the heterostructure \cite{Hu2020_ncomms, Wu:2020, Vidal2021}. For example, in the MnBi$_2$Te$_4$ bilayer, the excitonic and axion orders may couple, modulating the material's magnetoelectric response \cite{Liebman:2025}.
Overall, our result paves the way to engineering many-body states with dynamically controlled interplay of magnetism, topology and excitonic order.\\

\textit{During the final preparation of this manuscript we became aware of a related observation, based on trARPES experiments, of a potential exciton condensate state in monolayer films of MnBi$_2$Te$_4$ \cite{nguyen:2026}.}\\

\section{Experimental Section}
\textit{Crystal growth:} Single crystals were grown using the method previously reported in Ref.~\cite{Volckaert:2023}. The elemental starting materials of Mn, Sb, Bi, and Te were mixed and sealed in a vacuum quartz tube. The tube was subsequently heated to 900~$^{\circ}$C, slowly cooled to 601~$^{\circ}$C, and then canted to remove the flux.

\textit{Photoemission measurements:} The trARPES measurements were performed at the T-ReX facility (Trieste, Italy). A Ti:Sapphire Coherent Reg A laser system operated at 250~kHz with central energy of 1.55~eV was used to generate synchronized pump and probe pulses. The fourth harmonic (6.2~eV) of the fundamental beam was employed as an $s$-polarized probe. The fluence of the $p$-polarized pump pulses varied in the range of 20-120~\textmu J~cm$^{-2}$. The experimental time, energy and angular resolution were better than 200~fs, 50~meV and 0.2$\degree$, respectively. The surfaces of the samples were freshly prepared by cleaving \textit{in-situ} at a base pressure of $2\times10^{-10}$~mbar at room temperature. The sample temperature during measurement was 110~K unless otherwise stated.

\section{Acknowledgements}
We thank Simone Latini, Kristian S. Thygesen and Brian Moritz for insightful discussions.
Xing-Chen Pan would like to thank Bo Chen, Fucong Fei, and Fengqi Song for helpful discussions. The work was funded/co-funded by the European Union (ERC grant EXCITE with project number 101124619). Views and opinions expressed are however those of the author(s) only and do not necessarily reflect those of the European Union or the European Research Council. Neither the European Union nor the granting authority can be held responsible for them. The authors acknowledge funding from the Novo Nordisk Foundation (Project Grant NNF22OC0079960) and VILLUM FONDEN under the Villum Ascending Investigator Program (VIL83457) and the Villum Investigator Program (Grant. No. 25931). C.S. acknowledges Marie Sklodowska-Curie Postdoctoral Fellowship (project 101059528 MaPWave). Work at Advanced Institute for Materials Research has benefited from support of WPI-AIMR, JSPS KAKENHI Basic Science A (22H00278, 21K18590, 20H04623, 18H04473, 18H03858,  18F18328) and Tohoku University FRiD program.

\section{Author contributions}
P.M. and S.U. conceived the project. P.M., C.S., M.T., W.B., D.P. and F.C. performed the trARPES measurements. M.T., W.B., D.P. and F.C. maintained and prepared the trARPES facility for the experiments. X.-C.P. and Y.P.C. synthesized the crystals. P.M. analyzed the data and prepared the manuscript together with S.U.. All authors provided feedback on the final draft. 

\section{Conflict of interest}
The authors declare no conflict of interest.

\section{DATA AVAILABILITY}
The data used in this study is available on Zenodo  \cite{Zenodo}.

\clearpage

\section{Supplementary Note 1: Equilibrium characterization of the bulk valence band dispersion}

\begin{figure*}[h] 
\begin{center}
\includegraphics[width=\textwidth]{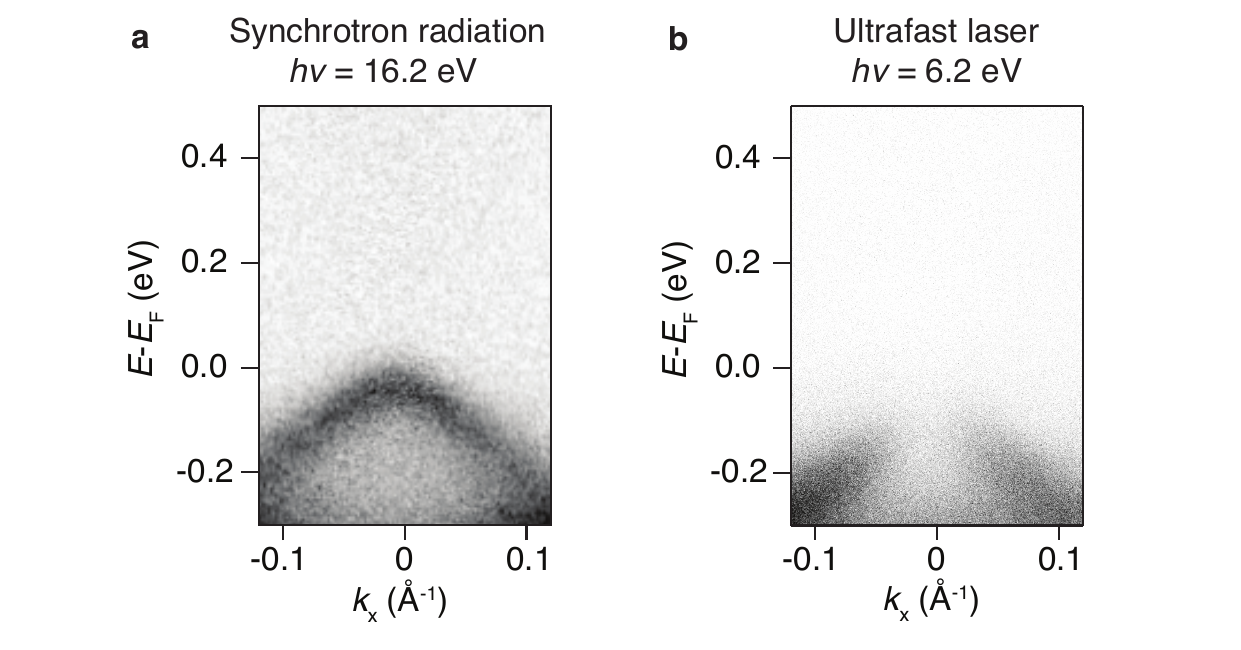}
\caption{(a) Equilibrium high-resolution ARPES spectrum of MnBi$_{1.7}$Sb$_{0.3}$Te$_4$ taken using synchrotron radiation at photon energy $h\nu = 16.2$~eV. (b) Equivalent ARPES cut taken with the ultrafast laser system at $h\nu = 6.2$~eV, without the pump beam present on the sample.}
\label{fig:s1}
\end{center}
\end{figure*}

In Figure~\ref{fig:s1}(a), we present a high-resolution equilibrium measurement of MnBi$_{1.7}$Sb$_{0.3}$Te$_4$ taken using synchrotron radiation ($h\nu = 16.2$~eV) at the SGM4 beamline, ASTRID2, Aarhus University \cite{jones2025spatial}. The measurement shows that the low-energy bulk valence band (BVB) dispersion is composed of a V-shaped band, and that at our doping level, the chemical potential is tuned into the bulk band gap.
Additionally, we show the equivalent measurement taken with the ultrafast laser probe ($h\nu = 6.2$~eV) in Fig.~\ref{fig:s1}(b). For this measurement, the pump beam was switched off such that the spectrum represents a true equilibrium situation. The intensity at the top of the BVB is suppressed at $h\nu = 6.2$~eV due to a photoemission matrix element effect.

\clearpage

\section{Supplementary Note 2: Energy distribution curve fits of bulk conduction band and exciton states}

\begin{figure*}[h] 
\begin{center}
\includegraphics[width=\textwidth]{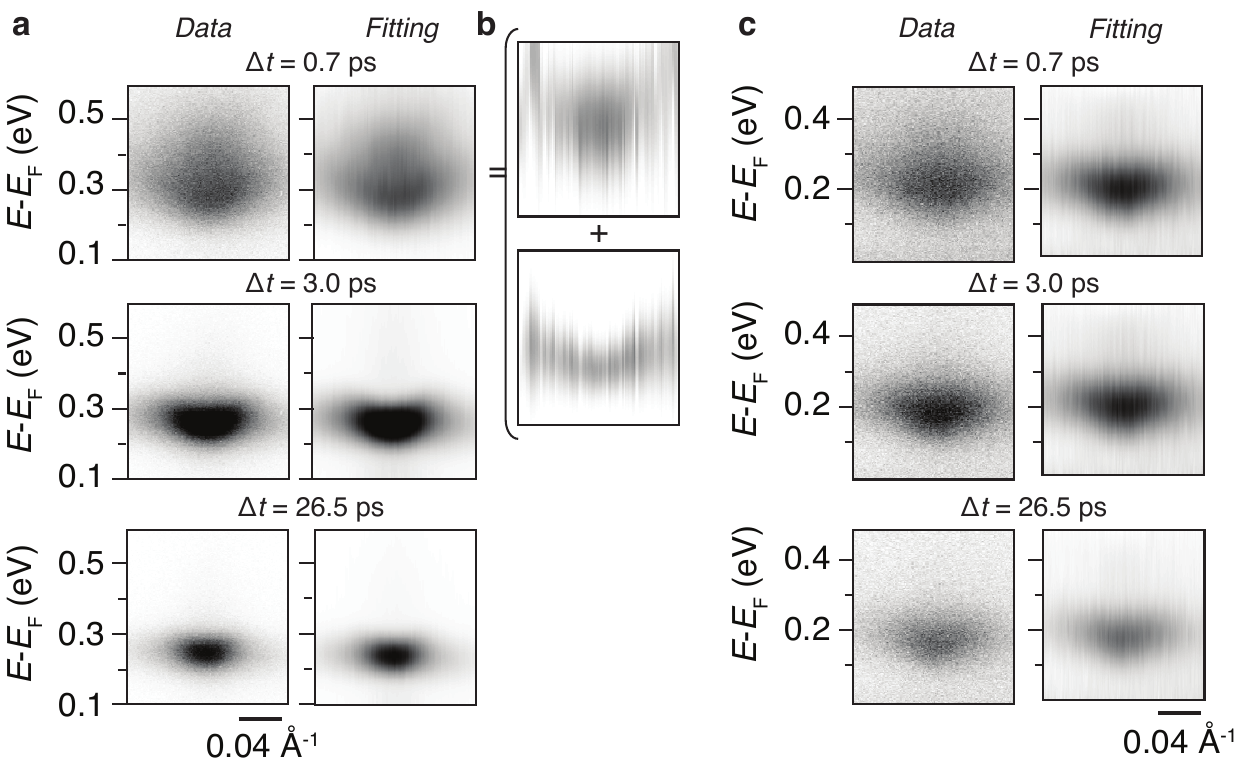}
\caption{(a), (c) Comparison between collected data (left) and EDC fit result (right) at selected time delays, for (a) $T = 110$~K and (c) room temperature. (b) EDC fit results for both peaks required to describe the spectrum at $\mathrm{\Delta} t = 0.7$~ps at $T = 110$~K.} 
\label{fig:s2}
\end{center}
\end{figure*}

Figure~\ref{fig:s2} presents the outcome of our energy distribution curve (EDC) fitting procedure in order to obtain the bulk conduction band (BCB) and exciton dispersion. The EDCs are fitted using a double or single Lorentzian function on a polynomial background across the momentum range of the bands. At $T = 110$~K, the spectra collected at early time delays ($\mathrm{\Delta} t < 3.00$~ps) require two peaks to describe the data, as shown in Figs.  \ref{fig:s2}(a)-(b). At room temperature, data at all time delays can be described by a single peak as shown in Fig.  \ref{fig:s2}(c). The quality of the EDC fits is illustrated in Fig.  \ref{fig:s2} by reconstructing the dispersion from the obtained fit parameters and comparing to the raw data.

\clearpage

\section{Supplementary Note 3: Analysis of the bulk valence band dynamics}

\begin{figure*}[h] 
\begin{center}
\includegraphics[width=0.9\textwidth]{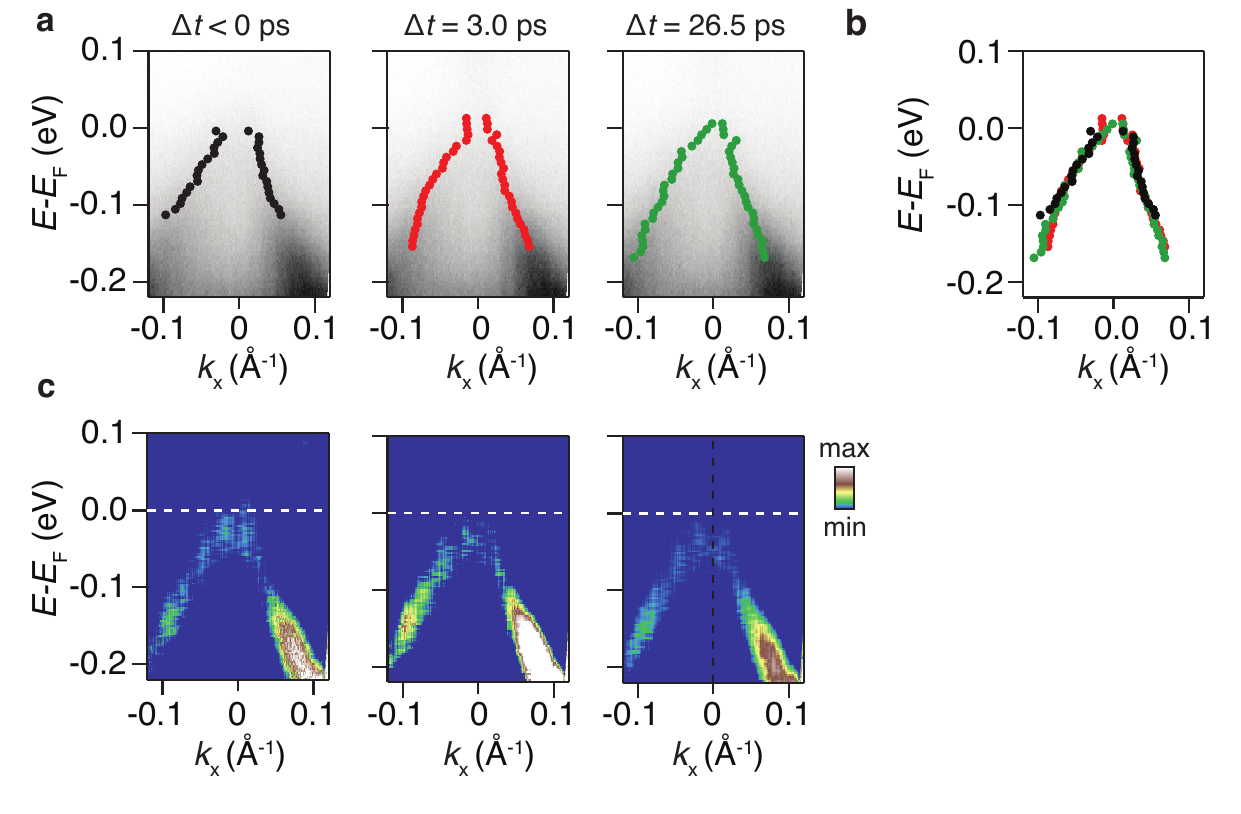}
\vspace{-0.8cm}
\caption{(a) Bulk valence band (BVB) region for the stated time delays. The overlaid markers are peak positions obtained by fitting a double Lorentzian function on a polynomial background to momentum-dependent curves (MDCs). (b) MDC fit results combined for the selected time delays in (a). (c) Second derivative images of the spectra in (a).}
\label{fig:s3}
\end{center}
\end{figure*}

We examine whether the BVB dispersion is modified upon optical excitation by fitting momentum distribution curves (MDCs) of the BVB to Lorentzian peaks. The fit results are overlaid on the ARPES intensity of the BVB at selected time delays in Fig.  \ref{fig:s3}(a). Comparing the extracted dispersion at each time delay in Fig.  \ref{fig:s3}(b) reveals that the BVB position remains fixed upon excitation. We can therefore exclude any influence of the surface photovoltage effect (SPV) on the long time dynamics we observe in the exciton state \cite{Hajlaoui2014, Ciocys2020} as this would manifest itself as a time-dependent rigid shift of all bands. 

The MDC fits become unreliable when approaching the top of the V-shaped band. However, a shift of the states located at the top of the BVB to lower energies can be seen via a change in the second derivative intensity extracted for each spectrum in Fig.  \ref{fig:s3}(c), suggesting a transient modification at the BVB edge.

\clearpage

\section{Supplementary Note 4: Evolution of the conduction band occupation at room temperature}

\begin{figure*}[h] 
\begin{center}
\includegraphics[width=\textwidth]{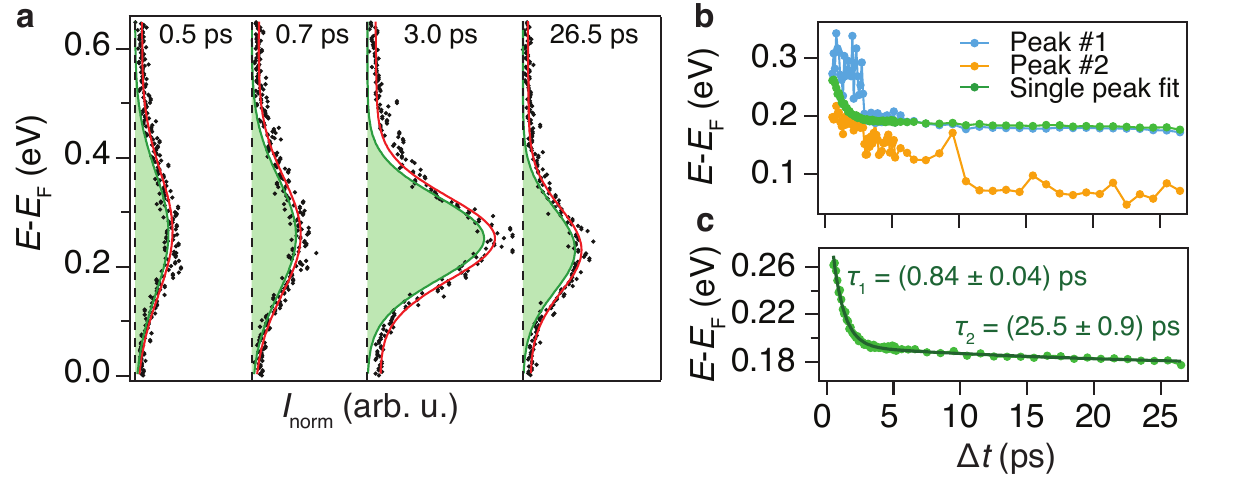}
\caption{(a) EDCs extracted at the \gp-point at selected time delays (black markers) fitted with a single Lorentzian function (red curve), for $T = 300$~K. Peak components are shaded in green. (b) Time-dependent peak position for two-peak fitting (blue and yellow traces) and
single-peak fitting (green trace). (c) Time-dependent peak position for single-peak model, with a double exponential fit overlaid (dark green curve).}
\label{fig:s4}
\end{center}
\end{figure*}

At room temperature, the dynamics of the excited state population is well-described by a single Lorentzian peak on a polynomial background, as seen in Fig.  \ref{fig:s4}(a). We tested both single- and two-peak fitting functions. However, for the two-peak model, the positions of the two peaks in Fig.  \ref{fig:s4}(b) are noisy and highly correlated, indicating overfitting. On the other hand, single-peak fitting in Fig.  \ref{fig:s4}(c) produces a stable time evolution of the position which exhibits double-exponential decay dynamics.

\end{document}